\documentclass[notitlepage,superscriptaddress,showpacs,nobalancelastpage,twocolumn,prl]{revtex4-1}
\usepackage{times}
\usepackage[english]{babel}
\RequirePackage[T1]{fontenc}
\RequirePackage{times}
\usepackage{amsfonts}
\usepackage{subfigure}
\usepackage{amsmath}
\usepackage{amssymb}
\usepackage{bm}
\usepackage{graphicx,epstopdf}
\usepackage{appendix}
\usepackage{array}
\usepackage{color}
\usepackage[hidelinks]{hyperref}
\usepackage{cancel}
\usepackage{my_symbols}
\usepackage[normalem]{ulem}
\usepackage{CJK}
\providecommand{\U}[1]{\protect\rule{.1in}{.1in}}

{\par}

\newcommand\rmv{\bgroup\markoverwith {\textcolor{red}{\rule[0.5ex]{2pt}{0.4pt}}}\ULon}

\begin{document}
\begin{CJK*}{UTF8}{gbsn} 
\title{Thermal induced monochromatic microwave generation in magnon-polariton}
\author{Vahram L. Grigoryan}
\affiliation{Institute for Quantum Science and Engineering, Southern University of Science and Technology, Shenzhen 518055, China}
\affiliation{The Center for Advanced Quantum Studies and Department of Physics, Beijing Normal University, Beijing 100875, China}
\author{Ke Xia}
\email[Corresponding author:~]{kexia@bnu.edu.cn}
\affiliation{Institute for Quantum Science and Engineering, Southern University of Science and Technology, Shenzhen 518055, China}
\affiliation{The Center for Advanced Quantum Studies and Department of Physics, Beijing Normal University, Beijing 100875, China}

\begin{abstract}
    We propose thermal induced generation of monochromatic microwave radiation in magnon-polariton. Mechanism of thermal to microwave energy transformation is based on intrinsic energy loss compensation of coupled magnon and microwave cavity oscillators by thermal induced "negative damping". A singularity at an exceptional point is achieved when the damping of the system is fully compensated at the critical value of "negative damping". At the exceptional point, the input energy is equally distributed between the magnon and photon subsystems of the magnon-polariton. The efficiency of transformation of thermal energy into useful microwave radiation is estimated to be as large as 17 percent due to magnon-photon coupling mediated direct conversation of spin current into microwave photons.
\end{abstract}
	\maketitle
\end{CJK*}


Magnon-polaritons are hybrid bosonic quasiparticles consisting of strongly interacting magnons and photons in microwave cavities \cite{dicke_1954,tavis_1968
}. Achieving the strong coupling regime triggered great interest in the field of cavity-spintronics  \cite{cao_2015,bai_2015,zhang_2017
}. Manipulation of spin current via magnon-photon coupling \cite{bai_2017}, indirect coupling between spin-like objects mediated by cavity \cite{rameshti_2018
}, and thermal control of the magnon-photon coupling \cite{castel_2017} are few of the effects. Despite the extensive study of magnon-polaritons, there is so far lack of considerable attention to an inherent dissipative nature of this systems.  Dissipating of the unbound magnon-polariton states leads to complex energies, which makes the Hamiltonian of the system non-Hermitian \cite{tannoudji_1977,moiseyev_2011}. In contrast to Hermitian systems, where off-diagonal coupling causes level energies to only repel each other and crossing is possible only for a vanishing interaction
, a striking property of non-Hermitian systems is that by tuning some parameters of the system, a singularity point in eigenfunctions and eigenvalues can be revealed \cite{heiss_2012}. The singularity point is called exceptional point (EP), where both real and imaginary energies can coalesce even for non vanishing interaction \cite{philipp_2000}.
Recently we proposed \cite{grigoryan_2018} a system where level attraction and EPs can be observed by introducing an additional phase controlled field driving the magnetization. The predicted level attraction have been studied experimentally \cite{note,bhoi_2019}. Similar feature due to the cavity Lenz effect has been reported recently \cite{harder_2018}. An experimental observation of the EP is reported in \cite{zhang_2017}, where an additional photon pump into the system provides gain to the coupled magnon-photon system. 


\begin{figure}[t!]
	\includegraphics[width=.8\columnwidth]{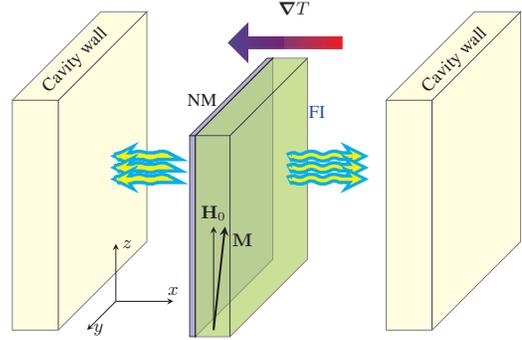}
	\caption{Schematic picture of the system with SSE-STT.}\label{fig:1}
\end{figure}

Here we propose a system, where 
microwave generation at the EP can be realized. Here, the energy loss is compensated via pumping energy into the magnon subsystem by means of "negative damping" or torque. Varying the "negative damping", level energy coalescence and EP occurs when the gain exactly compensates the losses of the system. We show that at the EP, the input energy is dissipated within the system itself resulting in dramatic microwave emission from the cavity. Moreover, in the presence of "negative damping", EP and microwave generation can be observed even in the absence of input field into the cavity. The "negative damping" can be applied via Spin Seebeck effect (SSE) induced spin transfer torque (STT) \cite{uchida_2010,holanda_2017,safranski_2017,note2
} or spin Hall effect (SHE)-STT \cite{chen_2016,hamadeh_2014,sklaner_2015}. Thus, the microwave emission at the EP in our proposal is a promising candidate for efficient transformation of electric/thermal input energy into microwave photon.

In \Figure{fig:1} we schematically illustrate the system, where the magnetic material with temperature gradient (for SSE-STT) is placed in a microwave cavity. We perform our calculations using semiclassical and scattering (see \textit{Supplementary Materials}) models, where the former provides intuitive and qualitative understanding confirmed by the latter model.
Our simple semiclassical picture is based on the combination of an effective LCR circuit for the photon in the cavity and Landau-Lifshitz-Gilbert (LLG) equations for spin \cite{bloembergen_1954,bai_2015,grigoryan_2018}. Two classical coupling mechanisms are Faraday induction of FMR \cite{silva_1999} and the magnetic field created by Ampere's law \cite{bai_2015}.

We consider a ferromagnetic insulator (FI) with the magnetization pointing in $\hzz$ direction due to crystal anisotropy, dipolar and external magnetic fields.
The effective LCR circuit for the cavity driven by a rf voltage is \cite{bloembergen_1954,bai_2015,grigoryan_2018}
\begin{align}
&L \dot{\bj} +R \bj +\smlb{1/ C}\int \bj dt=\bV^F \label{eq:LCR},
\end{align}
where $L,$ $C$, and $R$ are induction, capacitance, and resistance,  respectively, and $\bj=j\smlb{0,\cos\omega t,\sin\omega t}$ is the current oscillating in $\hyy$-$\hzz$ plane. The driving voltage $\bV^F$ is induced from precessing magnetization according to Faraday induction

\begin{figure*}[t!]
	\begin{tabular}{ccc}
		\includegraphics[width=.66\columnwidth]{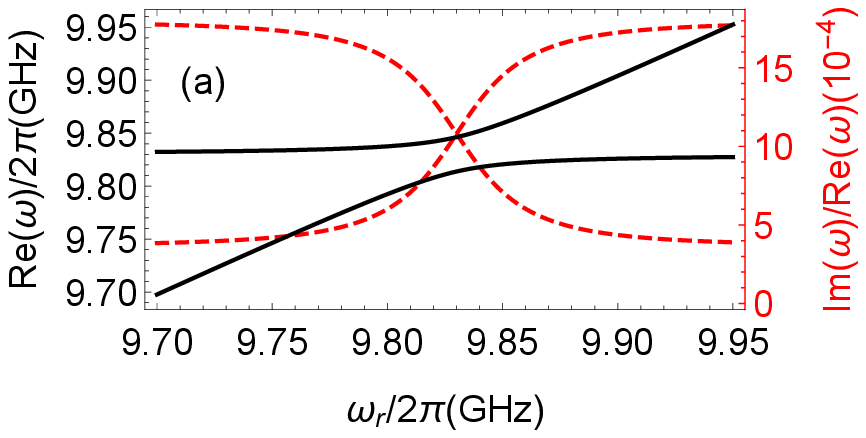}
		&\includegraphics[width=.66\columnwidth]{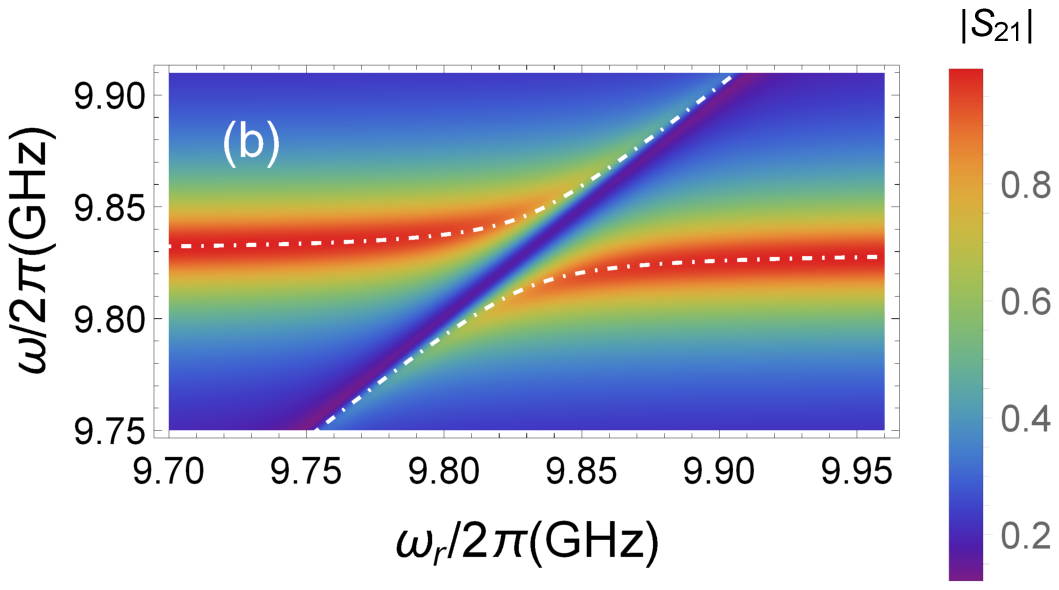}
			&\includegraphics[width=.62\columnwidth]{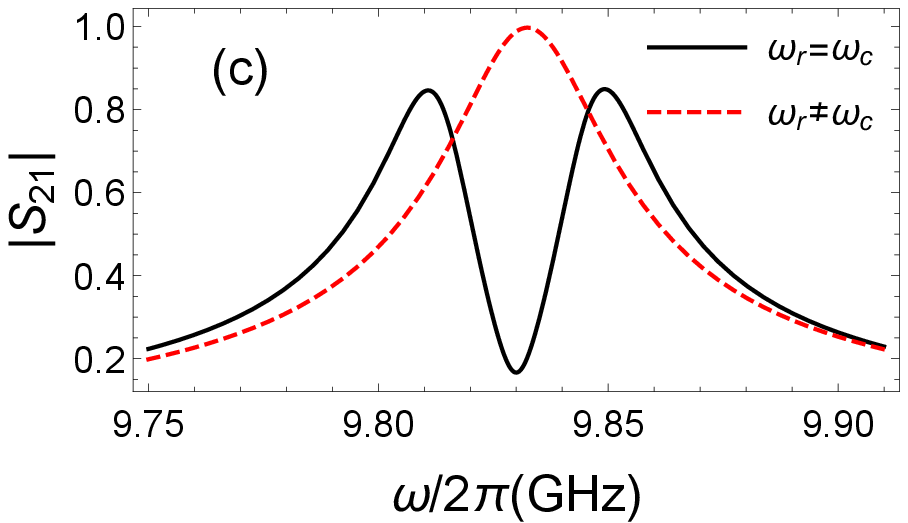}\\
			\includegraphics[width=.66\columnwidth]{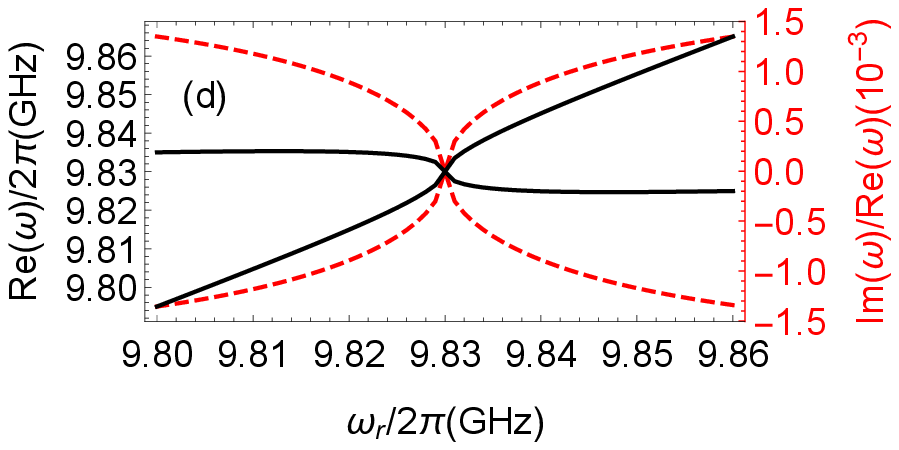}
&\includegraphics[width=.66\columnwidth]{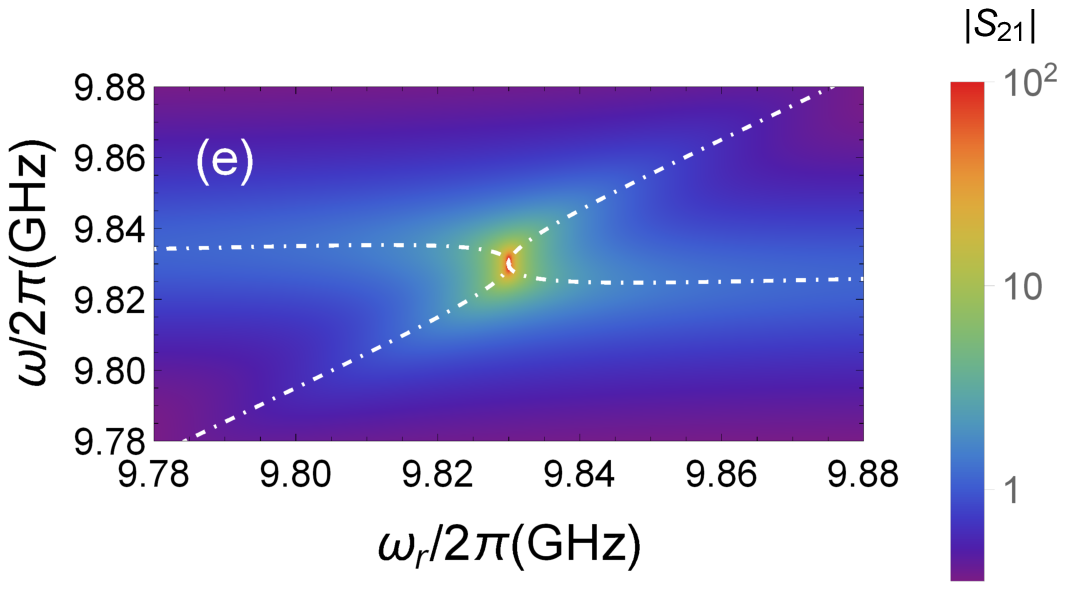}	&\includegraphics[width=.62\columnwidth]{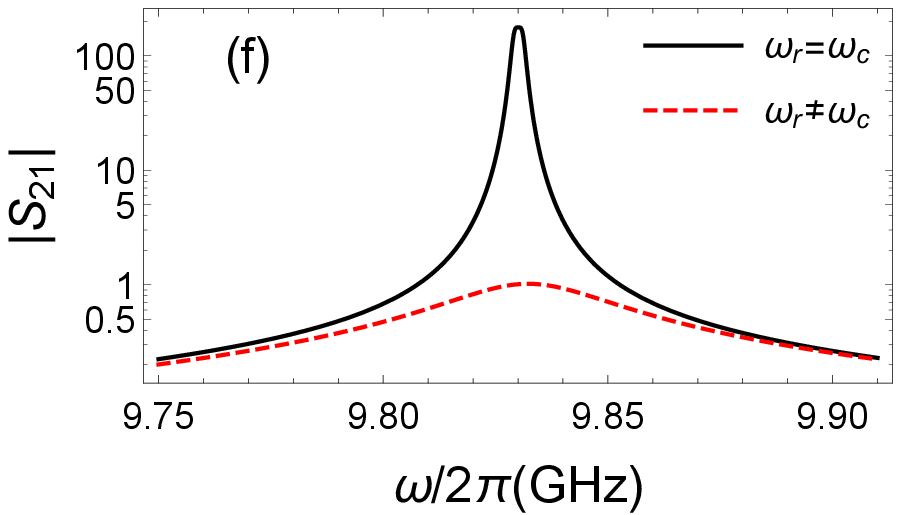}\\	
\includegraphics[width=.66\columnwidth]{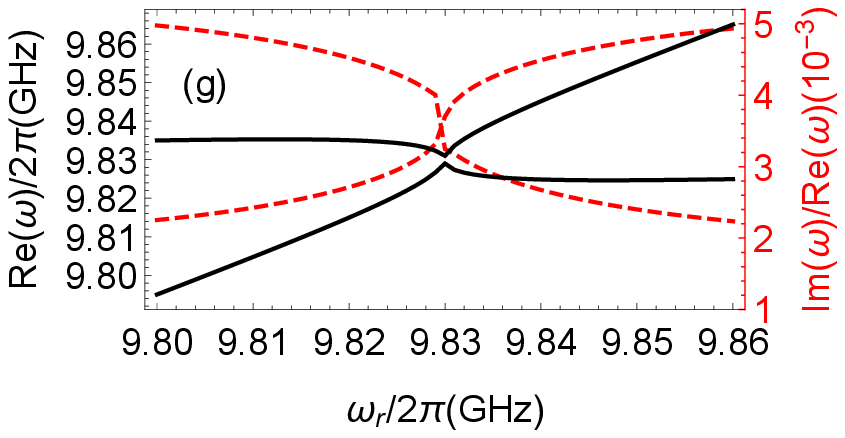}
	&	\includegraphics[width=.66\columnwidth]{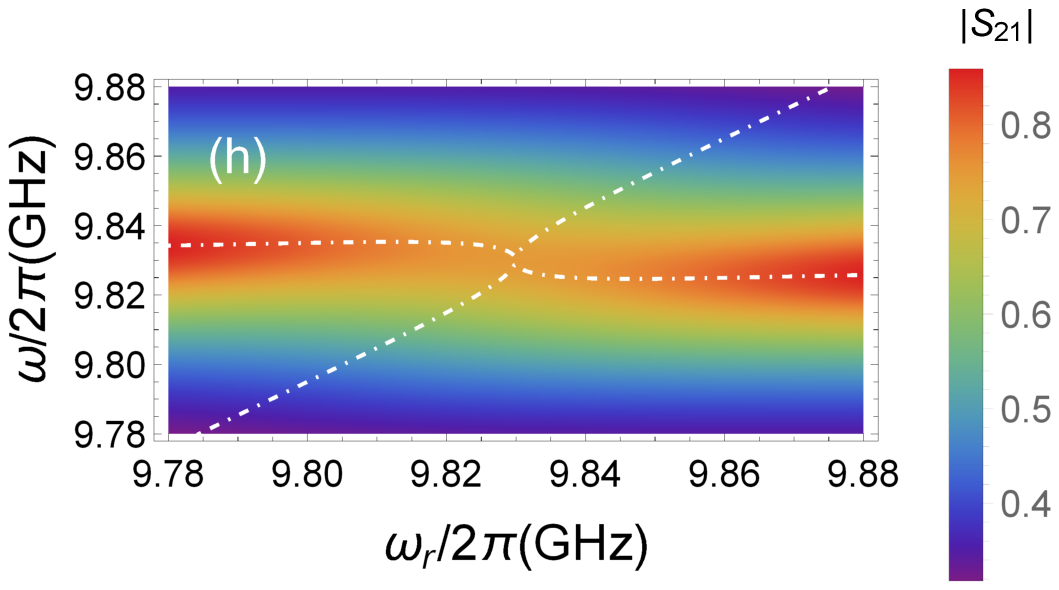}
&	\includegraphics[width=.62\columnwidth]{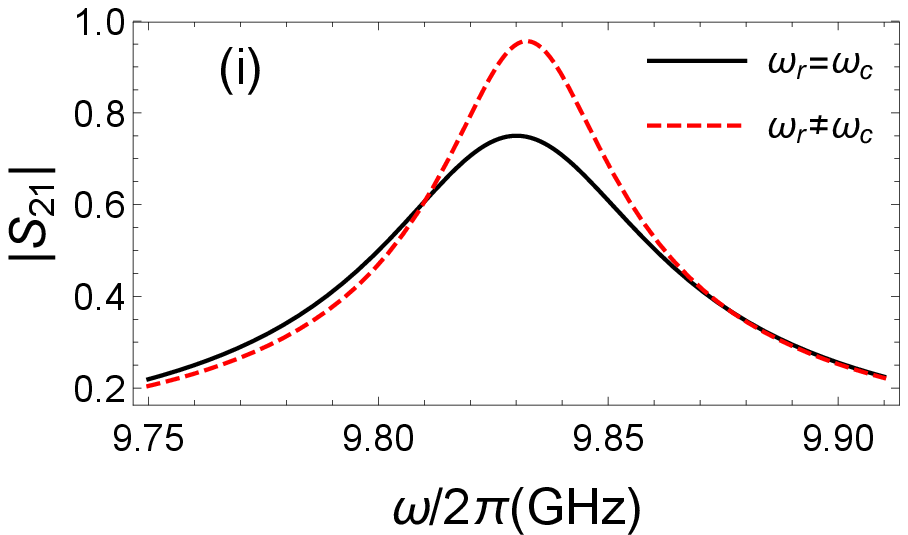}
	\end{tabular}\caption{(a) Real and imaginary components of mode energy levels as a function of FMR frequency, (b) transmission amplitude $\abs{S_{21}}$ as a function of $\omega$ and applied magnetic field calculated from semiclassical model \Eq{eq:transmission}, and (c) transmission amplitude as a function of $\omega$ at different FMR frequencies for $\omega_s=0.$ (d-f) The same as in (a-c) for $\omega_s=\omega_{s,EP} \simeq 0.0043 \omega_M.$ (g-i) The same as in (a-c) for $\omega_s\simeq -0.01\omega_M.$  }\label{fig:transmission2}
\end{figure*}

\begin{align}
&V^F_x\smlb{t}=K_c L\dot{m}_y,~V^F_y\smlb{t}=-K_c L\dot{m}_x\label{eq:2},
\end{align}
where $K_c$ is coupling parameter. The magnetization precession in the magnetic sample is governed by the LLG equation
\begin{align}
& \dot{\boldsymbol{M}}=-\gamma \mu_0 \boldsymbol{M}\times \boldsymbol{H}  + {\alpha \ov M_s} \boldsymbol{M}\times \dot{\boldsymbol{M}}+\boldsymbol{\tau}_\ssf{SSE}\label{eq:LLG},
\end{align}
where $\bM \simeq M_s \hzz+\mb,$ with $M_s$ being the saturation magnetization of FI, and $\mb=m$ $\smlb{\cos\omega t,\sin\omega t,0}$. $\gamma$ is gyromagnetic ratio and $\mu_0$ is the vacuum permeability. $\alpha$ is the intrinsic Gilbert damping parameter.  $\bH=\bH_{0}+\bh$ is the effective magnetic field in FI with $\bH_0=H_0\hzz$ being the sum of external magnetic, anisotropy, and dipolar fields aligned with $\hzz$ direction. $\bh$ is the induced magnetization from Ampere's law.
\begin{equation}
\bh=K_m \bj\times \hxx, \label{eq:aamp}
\end{equation}
where $\hxx$ is the wave propagation direction and $K_m$ is the coupling parameter. $\btau_\ssf{STT}=\smlb{\gamma \mu_0/M_s}$ $h_\ssf{STT}\bM\times \smlb{\bM\times \hzz}$  is the STT induced torque \cite{holanda_2017,yu_2017} (we neglect the effect of field-like torque \cite{zhu_2012}). The coupled LCR and LLG equations lead to
\begin{align}
&\Omega\smatrix{m\\h}=0 \qwith\nn
&\Omega=\smatrix{\omega-i \alpha \omega-\omega_r+i\omega_s& {\omega_M\ov 2}\\  {K^2 \omega^2\ov 2}  & \omega^2-2i \beta \omega \omega_c-\omega_c^2} ,\label{eq:mat}
\end{align}
where $\omega_M=\gamma \mu_0 M_s$ and $\omega_r \simeq \gamma \mu_0 H_0.$
$K\approx \sqrt{K_cK_m}.$ 
The cavity frequency $\omega_c=1/\sqrt{LC}$ and the cavity damping $\beta=R/\smlb{2L\omega_c}.$
$\omega_s\equiv \gamma \mu_0 h_\ssf{STT}$ can be tuned by e.g., the temperature gradient for SSE induced torque. By solving  \Eq{eq:mat} ($\det\Omega=0$) at given magnetic field we obtain roots for $\omega$. The two positive real components of $\omega$ determine the spectrum of the system, while imaginary parts describe damping.

We calculate the transmission amplitude using input-output formalism \cite{bai_2015,grigoryan_2018
}
\begin{align}
&\Omega  \smatrix{ m\\ h}=\smatrix{0\\ \omega ^2 h_{0}} \nn
&S_{21}=\Gamma h / h_{0}=\Gamma {\omega^2\smlb{\omega -i\alpha \omega -\omega_r+i\omega_s}\ov \det{\Omega}}
\label{eq:transmission},
\end{align}
where $\mu_0 h_{0}=10 \mu T$ \cite{cao_2015} is the input magnetic field driving the system, $\Gamma$ is a normalization parameter \cite{bai_2015}.


\begin{figure*}[t!]
	\begin{tabular}{ccc}
		\includegraphics[width=.7\columnwidth]{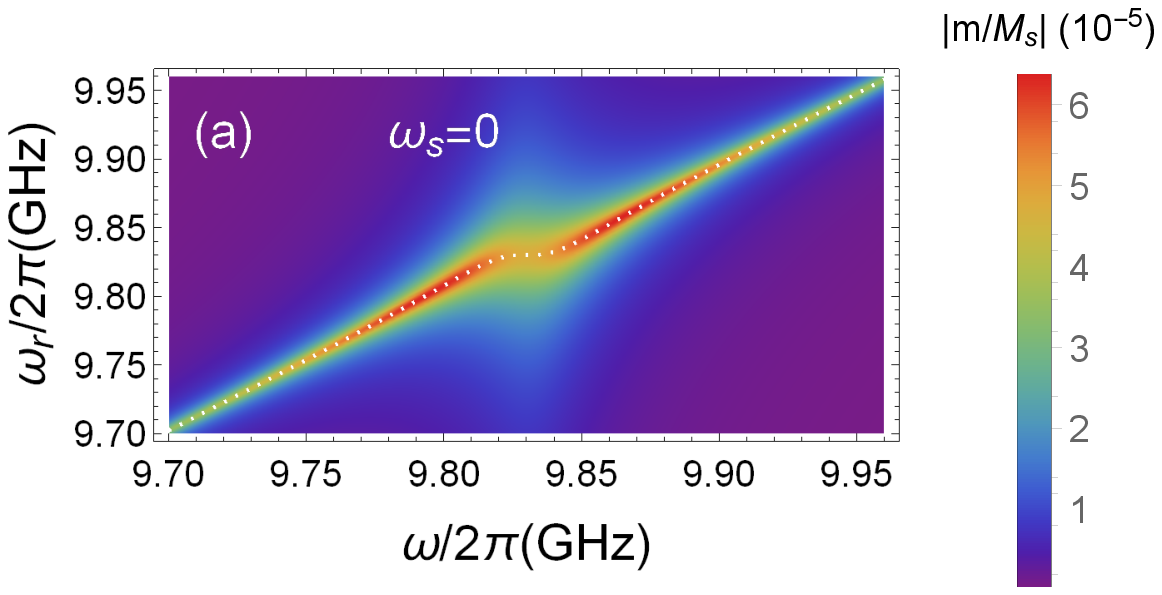}
		&\includegraphics[width=.68\columnwidth]{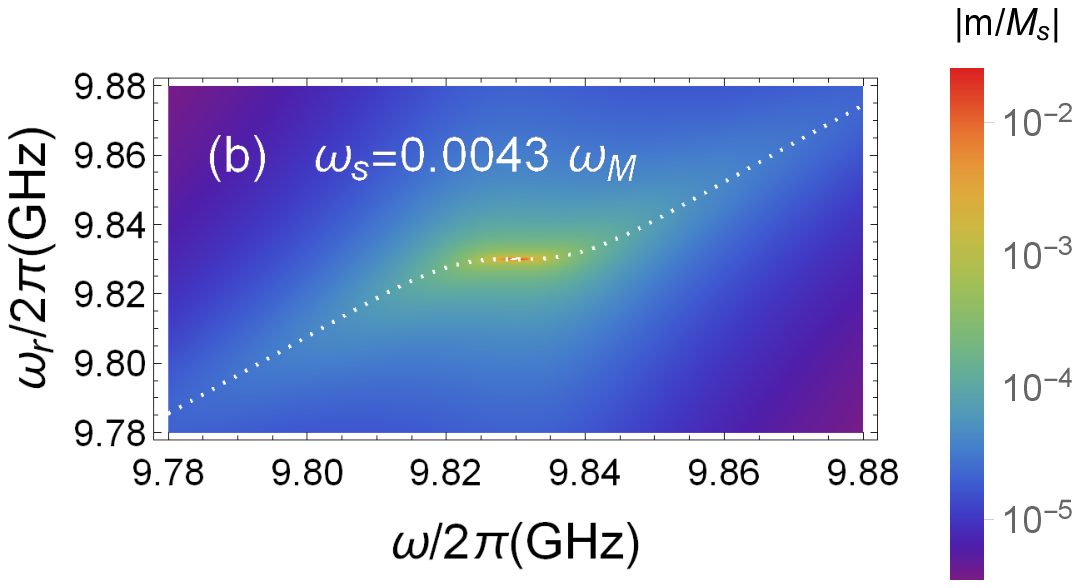}		&\includegraphics[width=.5\columnwidth]{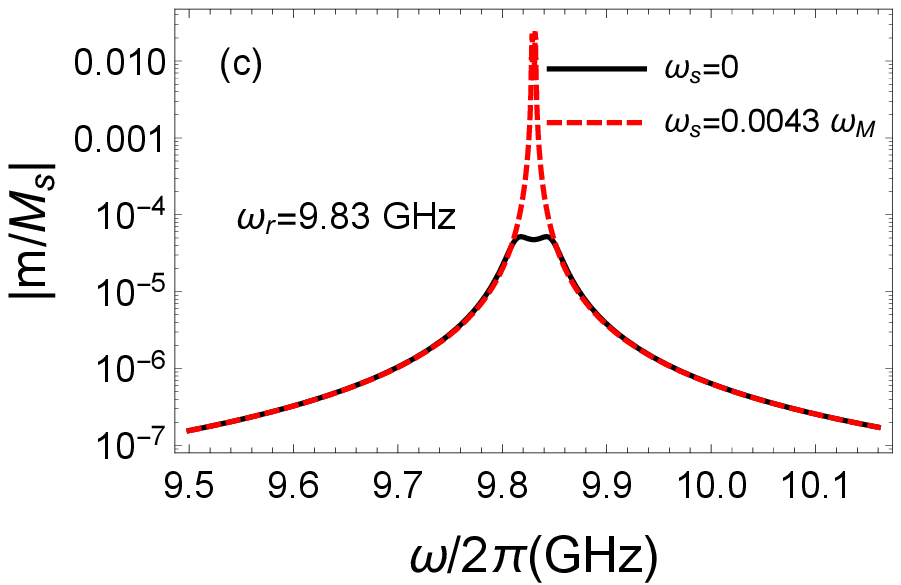}
	\end{tabular}\caption{FI magnetization as a function of $\omega$ and applied magnetic field (a) without "negative damping", and (b) with "negative damping" $\omega_g=\omega_{s,EP}.$ (c) Magnetization at resonant FMR frequency in the absence and presence of "negative damping" as a function of microwave frequency.}\label{fig:transmission_m}
\end{figure*}

\Figure{fig:transmission2} we show the spectrum and transmission amplitude for three different cases. To understand the spectrum and the transmission we analyse \Eq{eq:transmission}, where near the resonant point we have
\begin{equation}
\det\Omega=2 \omega_c^2 \beta \smlb{\omega_s -\alpha \omega_c -{K^2 \omega_M\ov 8 \beta}}. \label{eq:detom}
\end{equation} and the spectrum calculated from $\det\Omega=0$ can be simplified as
\begin{equation}
\omega=\frac{1}{2} \left\{  \omega_r+\omega_c+i\left( \alpha \omega_c +\beta\omega_c -\omega_s  \right) \pm \omega_g  \right\}  \label{eq:spgen}
\end{equation}
with the Rabi frequency being
\begin{equation}
\omega_g=\sqrt{\left[\left(\omega_r-\omega_c\right)+ i \left(\alpha \omega_c-\beta\omega_c-\omega_s \right) \right]^2+\frac{K^2 \omega_M \omega_c}{2}} \label{eq:rabifreq}.
\end{equation}
In the absence of "negative damping" ($\omega_s=0$), from Eq. (\ref{eq:rabifreq}) we can recover the Rabi gap expression at resonance ($\omega_r=\omega_c$) in strong coupling regime ($K>\alpha,\beta$)  $\omega_g\approx K \sqrt{\omega_M\omega_c /2}$ \cite{bai_2015}. It also follows that in this regime the imaginary parts of $\omega$ are equal for two hybridized modes. In \Figure{fig:transmission2} (a-c) we show the results for this conventional coupling system. \Figure{fig:transmission2} (a) is the usual anticrossing of hybridized level energy modes ($Re (\omega)$) with $2\omega_g$ gap between the modes and crossing of dampings ($Im(\omega)$) at resonant cavity frequency $\omega_c/2\pi=9.83$ GHz \cite{bai_2015}. Corresponding transmission amplitude of the system is shown in \Figure{fig:transmission2} (b), where the coloured area shows the transmission and the spectrum is presented by dashed lines. The characteristic two peak behaviour of the transmission is shown by solid line in \Figure{fig:transmission2} (c) when ferromagnetic resonance (FMR) frequency matches with the cavity mode frequency ($\omega_r=\omega_c$). The dashed line is the out-of-resonant ($\omega_r\neq \omega_c$) transmission, which has a single peak at $\omega=\omega_c.$

Next, we analyse the possibility to encircle EP in strong coupling regime. It follows from Eq. (\ref{eq:spgen}) that to have coalescence of real and imaginary components of two hybridized modes one has to require
\begin{equation}
\left.\omega_g\right|_{\omega_r=\omega_c}=0. \label{eq:cond1}
\end{equation}
Plugging solution of Eq. (\ref{eq:cond1}) into Eq. (\ref{eq:detom}) and tuning either the positive coupling strength (EP for negative effective coupling has been discussed elsewhere \cite{grigoryan_2018}) or the damping of cavity \cite{zhang_2017}, the second condition $\det\Omega=0$ can  be fulfilled
\begin{equation}
\omega_s -\alpha \omega_c -K^2 \omega_M/ \left( 8 \beta\right)=0 \label{eq:cond2}.
\end{equation}
The critical value for the "negative damping" satisfying Eqs. (\ref{eq:cond1}) and (\ref{eq:cond2}) is $\omega_{s,EP}=\left(\alpha+\beta\right)\omega_c$ when the coupling equals to $K=2\beta \sqrt{2\omega_c}/\sqrt{\omega_M}.$ It follows from Eq. (\ref{eq:transmission}) that the second condition leads to singularity and dramatic photon emission near the EP. From Eq. (\ref{eq:spgen}), the critical value for reaching the emission peak corresponds to full compensation of magnetization and cavity dampings, $\im{\omega}=0$. For the parameters $\alpha=3.6 \times 10^{-4},$ $\beta\simeq 1.8 \times 10^{-3},$  $K=0.0072$ \cite{bai_2015}, $\omega_M/(2\pi)=\gamma \mu_0 M_s= 4.9 $ GHz, where $\gamma/(2\pi)=28 $ GHz/T, and $M_s \mu_0=0.175$ T \cite{cao_2015}, the critical value of the "negative damping" satisfying the conditions of EP is $\omega_{s,EP}=\left(\alpha+\beta\right)\omega_c\simeq 0.0043 \omega_M.$ In \Figure{fig:transmission2} (d) we plot the spectrum when both Eqs. (\ref{eq:cond1}) and (\ref{eq:cond2}) are satisfied. Both, real (solid curve) and imaginary (dashed curve) components of $\omega$ coalesce at the resonant frequency. Moreover, imaginary component equals to zero at the EP. The resulting transmission is shown in  \Figure{fig:transmission2} (e), where the bright spot encircles the EP. A large photon emission at the bright spot is shown in \Figure{fig:transmission2} (f), where the dashed and solid lines are the out-of-resonant and resonant (at EP) transmissions, respectively.


\begin{figure}[b!]
	\begin{tabular}{cc}
		\includegraphics[width=0.45\columnwidth]{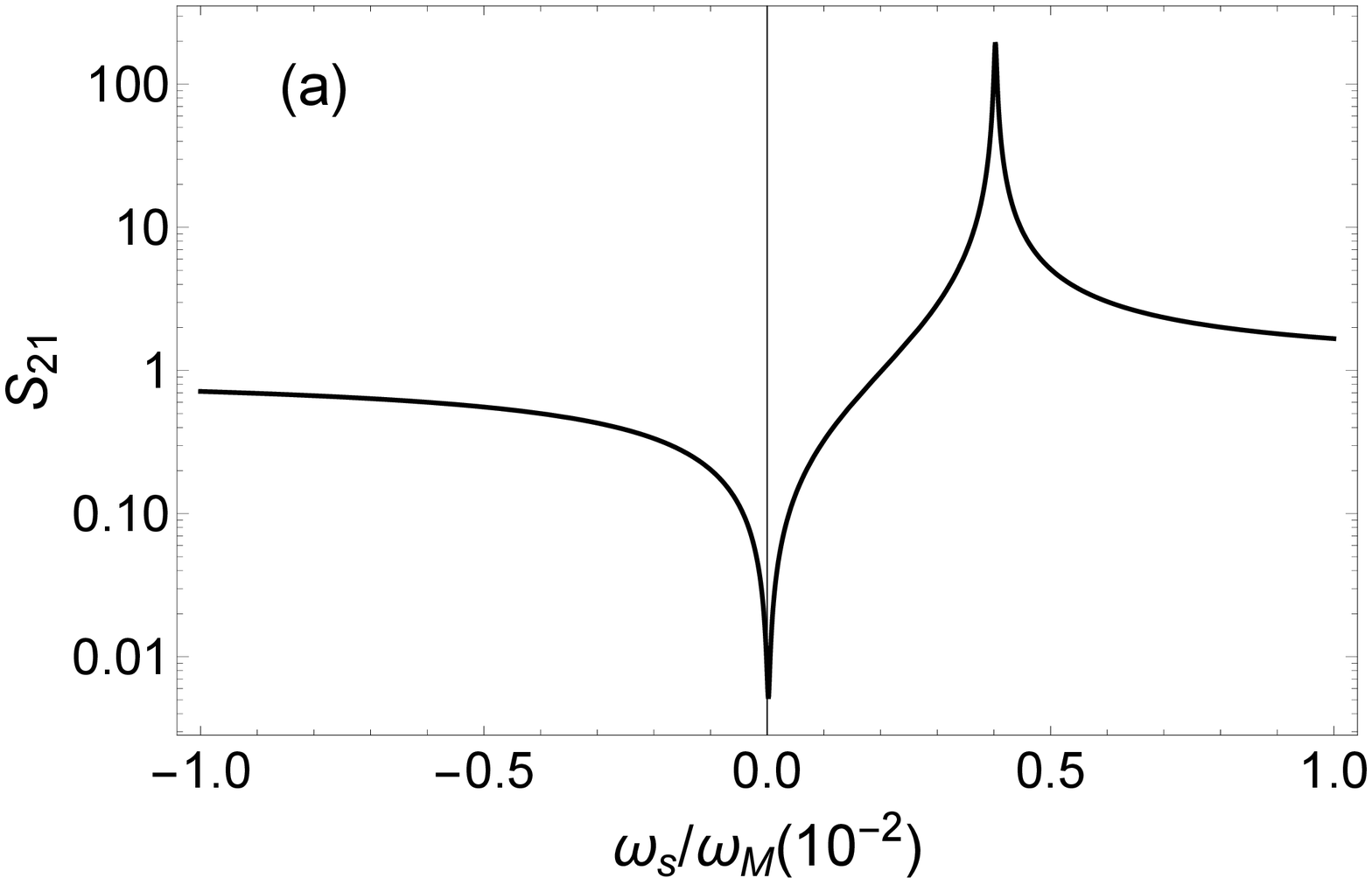}
		&	\includegraphics[width=0.45\columnwidth]{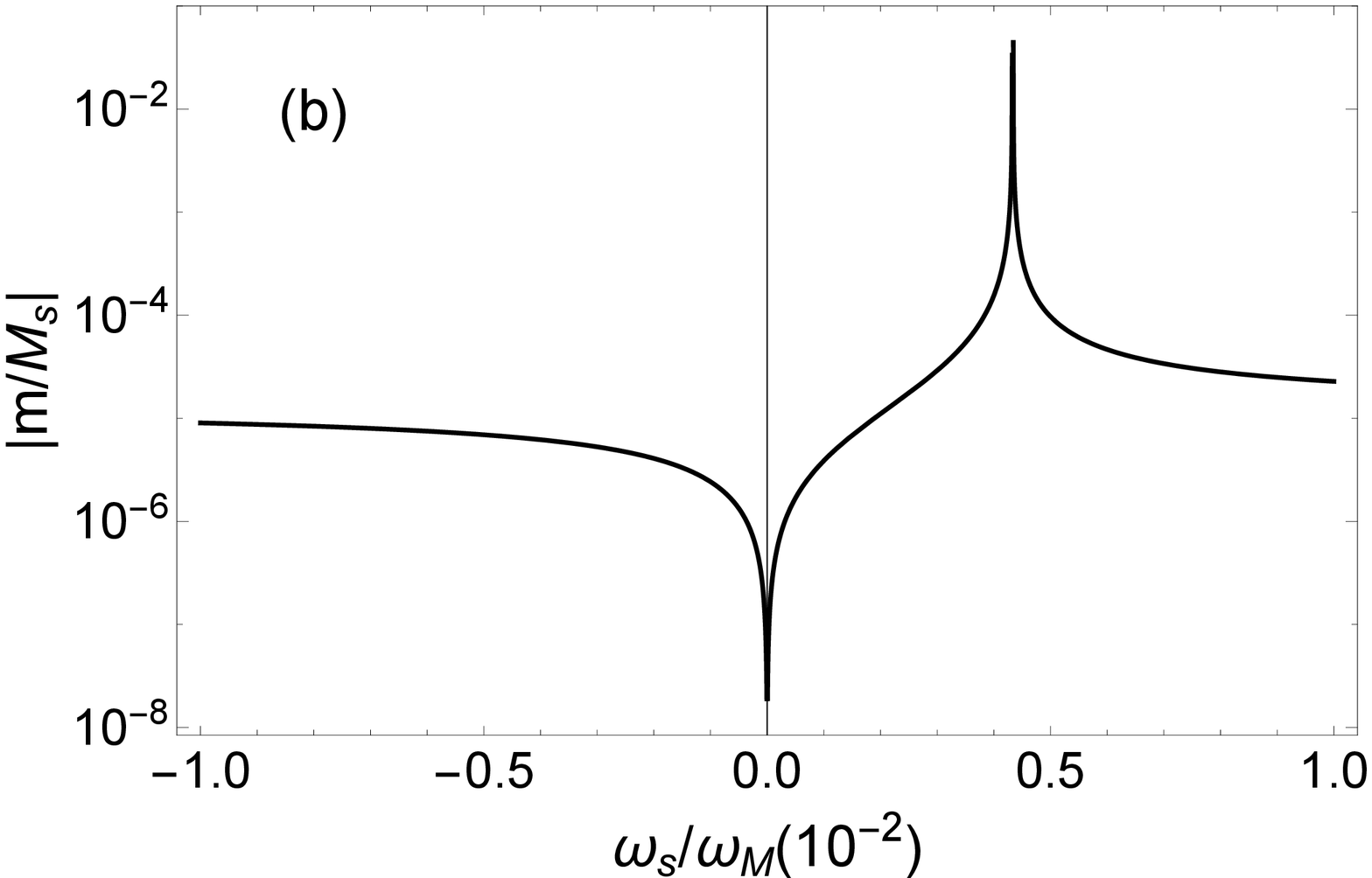}
	\end{tabular}
	\caption{Transmission amplitude (a) and the FI magnetization (b) as a function of STT $\omega_s.$ }\label{fig:tronoms}
\end{figure}

For a positive coupling, Eq. (\ref{eq:cond1}) has second solution for $\omega_s$ at which the real and imaginary components of level energy modes coalesce. The negative solution $\omega_s=\left(\alpha-3\beta\right)\omega_c \simeq -0.01 \omega_M$ corresponds to positive damping of the magnetization, which, together with intrinsic Gilbert damping, transform the system into effective weak coupling regime, when the coupling is smaller than the effective enhanced damping of the system. In \Figure{fig:transmission2} (g) we show the spectrum for second EP. Although the condition Eq. (\ref{eq:cond1}) is satisfied, the second condition in Eq. (\ref{eq:cond2}) is not fulfilled. As a results, complex energy levels coalescence at resonance, but, as shown in \Figure{fig:transmission2} (h) and (i), there is no large photon emission from the cavity due to non-zero damping of the system.


\begin{figure}[t!]
	\begin{tabular}{cc}
		\includegraphics[width=.45\columnwidth]{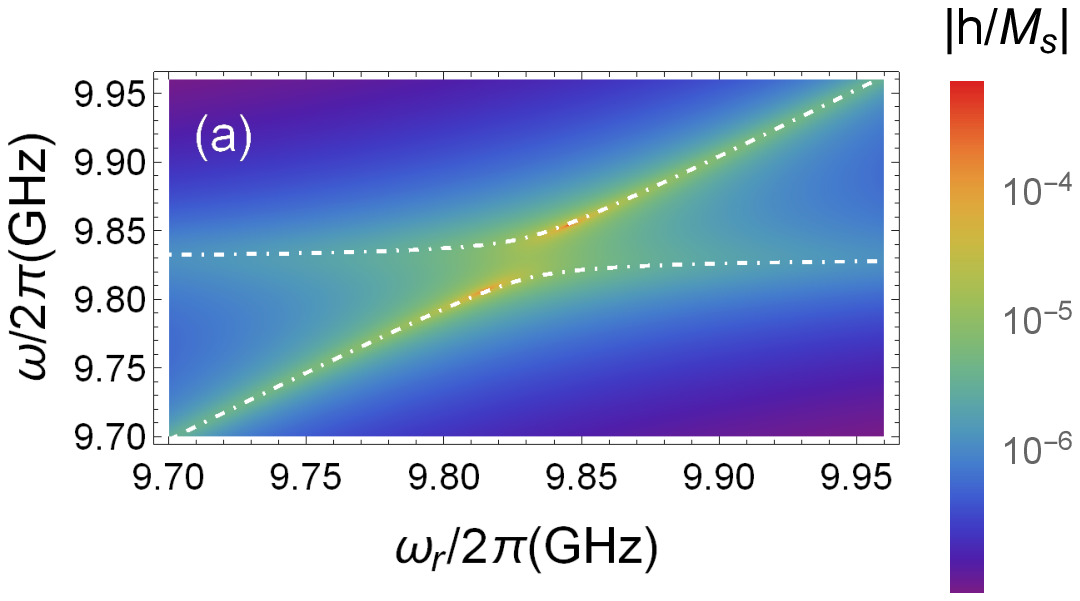}
&\includegraphics[width=.4\columnwidth]{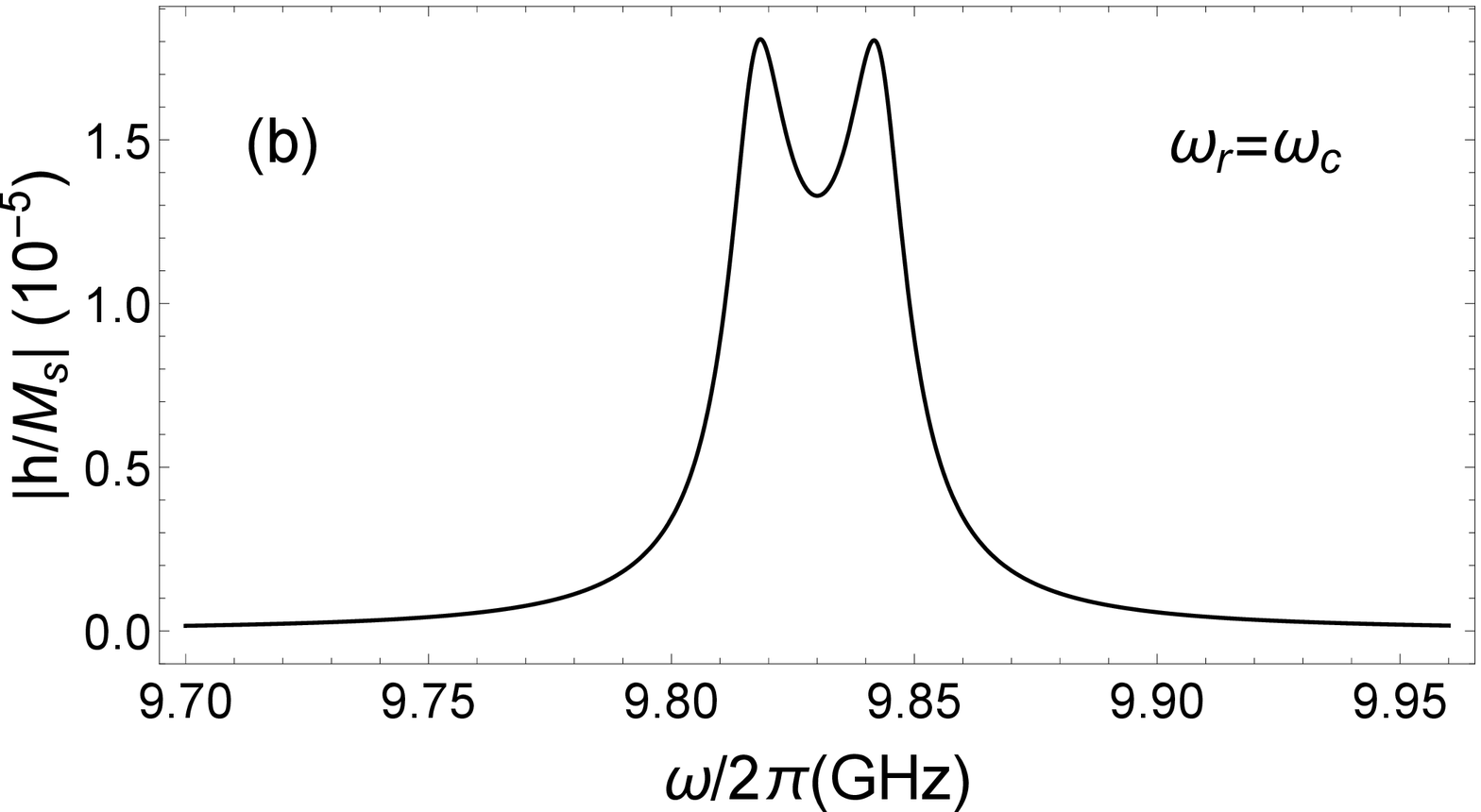}\\
		\includegraphics[width=.45\columnwidth]{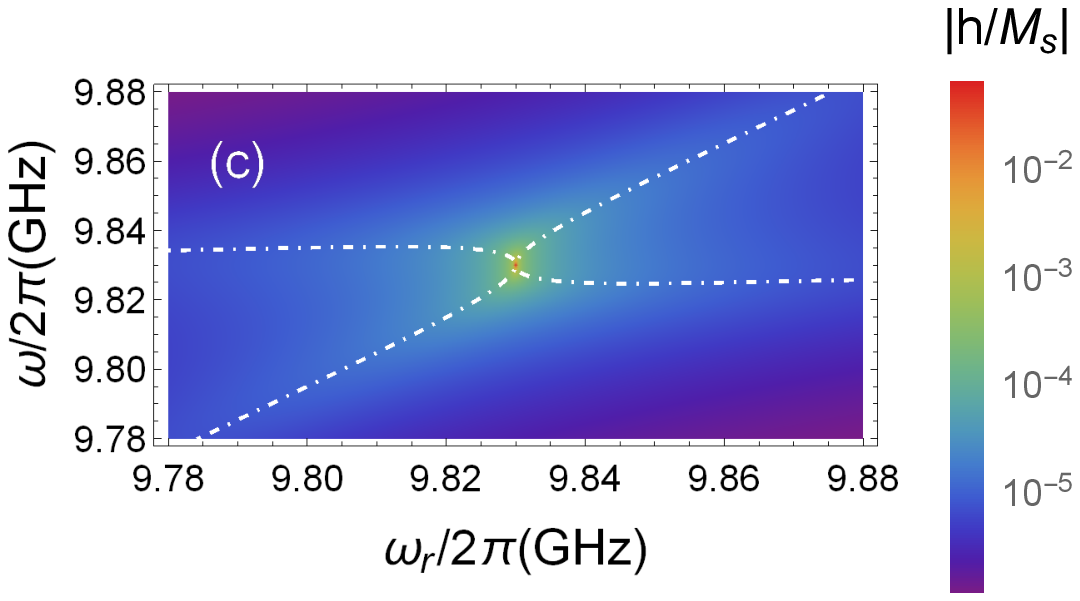}&
		\includegraphics[width=.4\columnwidth]{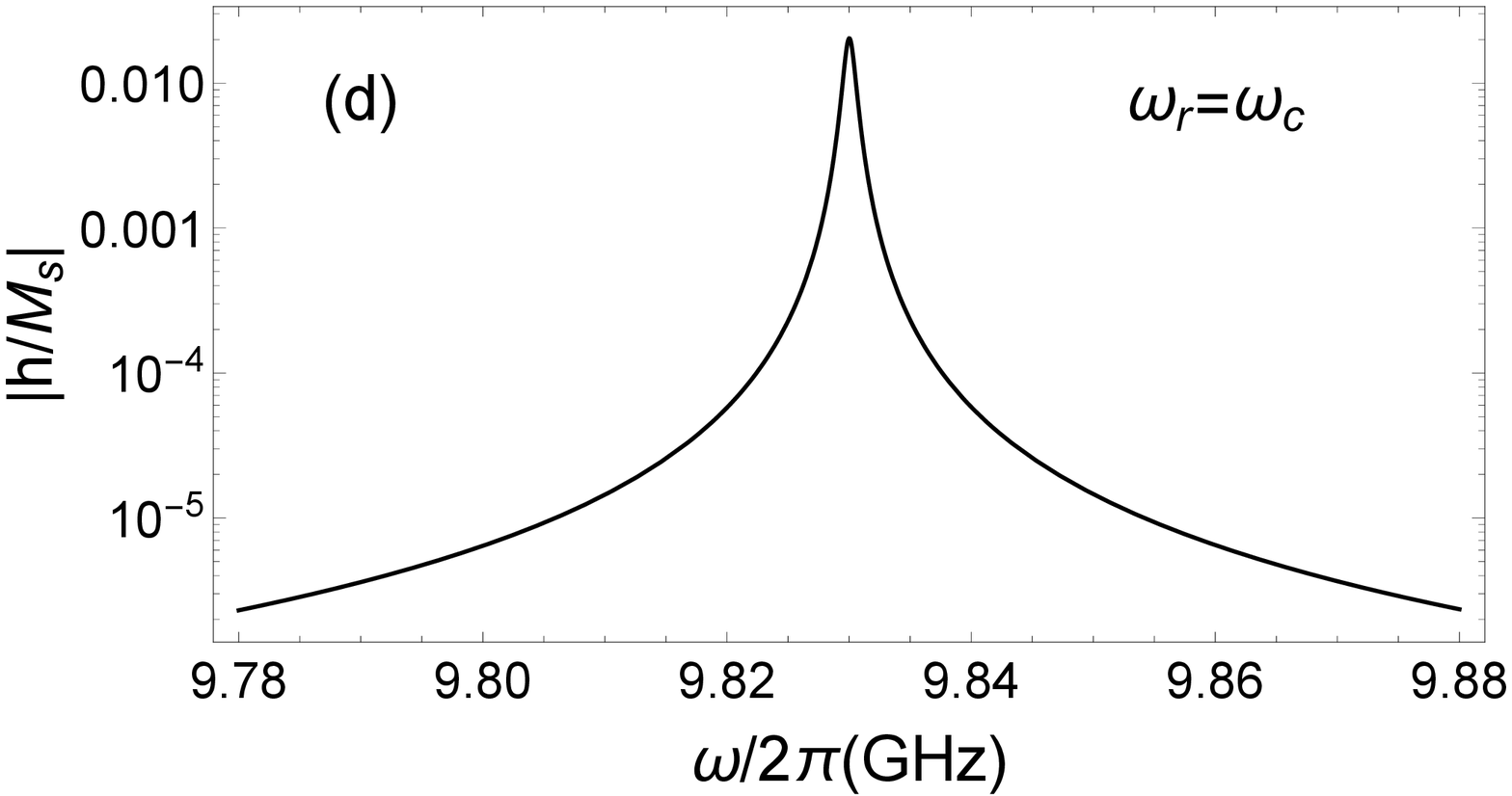}
	\end{tabular}\caption{(a) Output microwave amplitude $\abs{h/m_0}$ as a function of $\omega$ and applied magnetic field calculated from \Eq{eq:transmission1}, and (b)  the output as a function of $\omega$ at different FMR frequencies for $\omega_s=0.$ (c-d) The same as in (a-b) for $\omega_s=\omega_{s,EP} \simeq 0.0043 \omega_M.$ }\label{fig:transmission}
\end{figure}

Output magnetic field calculated from Eq. (\ref{eq:transmission}) provides understanding of the "negative damping" effect on the microwave photons in the cavity. To explore the effect of STT on the magnetic subsystem, we focus on the magnetization $m$ from Eq. (\ref{eq:transmission}). Using the same parameters as that for $S_{21},$ we plot in \Figure{fig:transmission_m} (a) the evolution of magnetization as a function of $\omega$ and $\omega_r$ in the absence of "negative damping". The dotted curve is the real component of the single solution for $\omega_r$ from $\det\Omega=0$ in Eq. (\ref{eq:mat}) \cite{bai_2015}. \Figure{fig:transmission_m} (b) shows the same when now the "negative damping" is at critical value, $\omega_s=\omega_{s,EP}$. Comparison of the magnetization values at resonant FMR frequency for this two cases is shown in \Figure{fig:transmission_m} (c). It is seen, that similar to $S_{21},$ the magnetization too, shows dramatic enhancement at the EP. In \Figure{fig:tronoms} (a) and (b), we plot the dependence of the transmission amplitude and the magnetization on $\omega_s$ at resonant FMR frequency, respectively. For SSE-STT, the absence of temperature gradient ($\omega_s=0$) leads to minimal value of transmission and magnetization at FMR frequency. This is because at resonant magnetic field the FMR and cavity modes are coupled and the cavity mode transmission (\Figure{fig:transmission2} (b)) and FI magnetization (\Figure{fig:transmission_m} (a)) split into two peaks and are suppressed at the anticrossing point \cite{cao_2015}. Negative value of $\omega_s,$ corresponds to spin current flowing from FI to normal metal (spin pumping) \cite{holanda_2017}, resulting damping enhancement. In the opposite case, when $\omega_s$ is positive, STT acts as "negative damping" and leads to large output in both subsystems, when the torque compensates the energy loss of the system.

It is known that spin torque can not only decrease the intrinsic damping but can drive magnetic oscillations in the absence of external microwave field \cite{collet_2016,safranski_2017,demidov_2017
}.  If such spin-torque oscillator (STO) is placed in the cavity, the oscillating magnetic field induces electric field which, in turn, creates magnetization due to Ampere's law \cite{bai_2015}. The induced field then acts back on the oscillator's magnetization. Here we show that due to inherent coupling between magnetization oscillations and the cavity field, the effects described in previous sections survive even in the absence of input microwave field. Without input field we write Eq. (\ref{eq:transmission})
\begin{equation}
\Omega  \smatrix{ m\\ h}=\smatrix{i \omega_s m_0\\ 0}
\label{eq:transmission1},
\end{equation}
where $m_0\simeq 1.54 \times 10^{-5}M_s$ (calculated for $\omega=\omega_r \neq \omega_c$) is the initial transverse magnetization. In \Figure{fig:transmission} (a) we plot the output signal as a function of microwave and FMR frequencies in the absence of input microwave field for "negative damping" value smaller that the critical value ($\omega_g\simeq 0.002\omega_M<\omega_{s,EP}).$ Near resonant frequencies, the spectrum behaves as usual coupled system, except that output signal occurs solely at near resonant frequencies. The reason is that when FMR frequencies are far from cavity mode frequency, the induced microwave is being absorbed by the cavity. \Figure{fig:transmission} (b) shows the output as a function of $\omega$ at resonant FMR frequency with anticrossing behaviour. The output signal at $\omega_{s,EP}$ is shown in \Figure{fig:transmission} (c-d). It is seen from \Figure{fig:transmission2} (e,f) \Figure{fig:transmission} (c,d) that the existence of large output signal does not depend on input microwave field into the cavity.
\begin{figure}[t!]
	\begin{tabular}{cc}
		\includegraphics[width=.55\columnwidth]{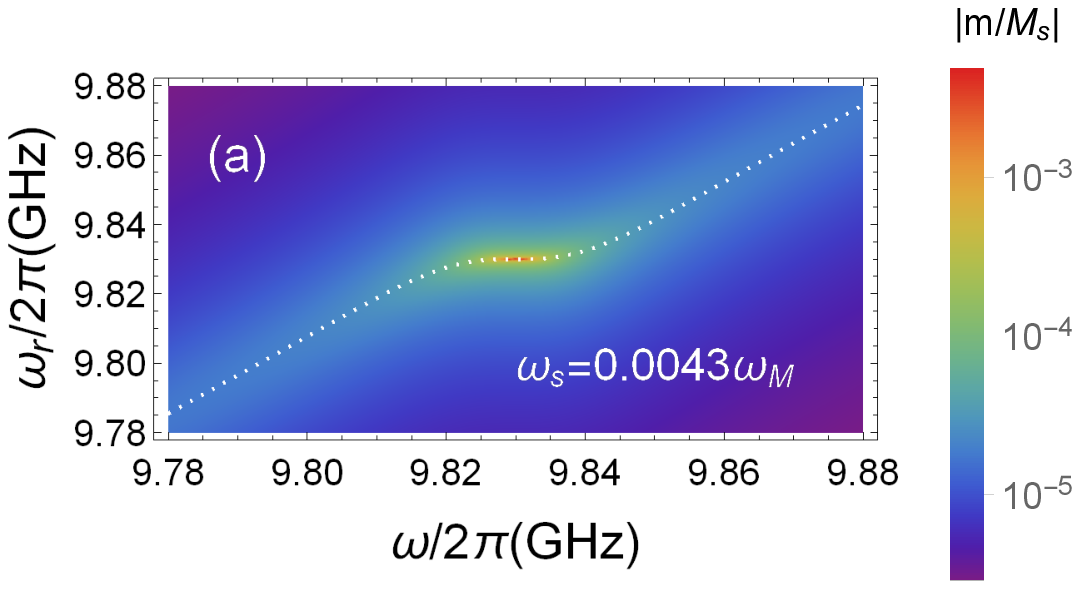}
		&\includegraphics[width=.4\columnwidth]{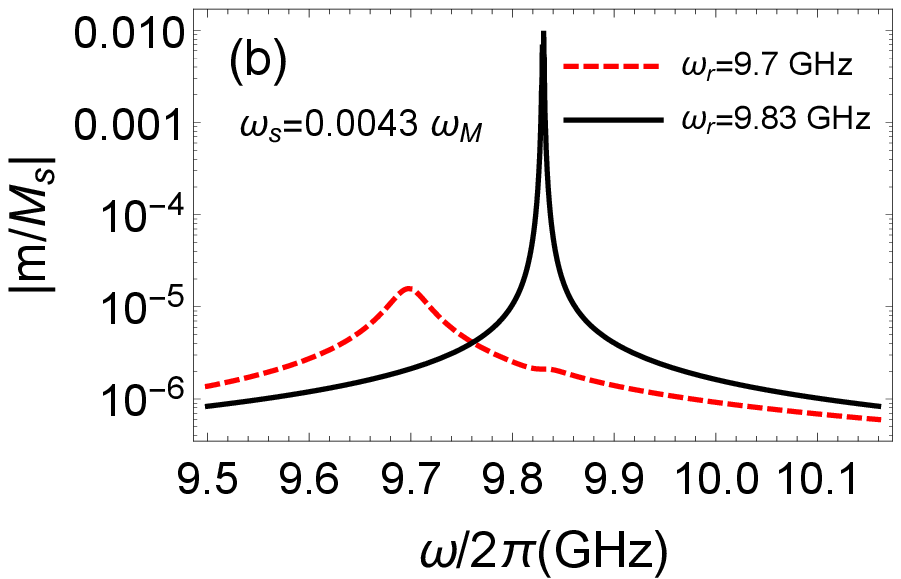}
	\end{tabular}\caption{(a) FI magnetization as a function of $\omega$ and applied magnetic field with "negative damping" $\omega_g=\omega_{s,EP}$ in the absence of input microwave field. (b) Magnetization at resonant and out-of-resonant FMR frequencies. }\label{fig:transmission11}
\end{figure}
This feature indicates that if the "negative damping" is sufficient to compensate the intrinsic damping of the magnon-polariton, the system can also be utilized for transformation of thermal (or electric energy in case of SHE-STT) energy into microwave. The magnetization evolution in the absence of input microwave field is shown in \Figure{fig:transmission11} (a). The comparison of the magnetization at resonant and out-of-resonant FMR frequencies in \Figure{fig:transmission11} (b) demonstrates the similar enhancement of the magnetization due to "negative damping." Due to energy loss compensation at the EP, the input energy by STT is dissipated within the system itself \cite{zanotto_2014} and distributed between the magnetic and photon subsystems, $P_{Tot}=P_{ph}+P_{m},$ where $P_{ph}\propto \abs{h}^2$ and $P_{m}\propto \abs{m}^2$ are the output powers of photon and magnon subsystems, respectively . We calculate the power distribution at the EP to be $P_{ph}/P_{Tot}=P_{m}/P_{Tot}=0.5$ meaning that the input energy is equally distributed between the subsystems. We estimate power conversion efficiency (ratio of useful output and input powers) in our system (see Supplementary Materials for details) to be $\eta/\eta_c=ZT/2(ZT+1)\approx 0.17,$ where figure of merit is $ZT\approx 0.53.$ Relatively large value of the efficiency compared to SSE induced thermoelectric devices \cite{uchida_2016} is due to direct conversion of magnon current into useful microwave power by avoiding spin current injection into adjacent normal metal and spin to charge conversion by inverse spin Hall effect.

For estimation of thermal gradient induced effects we use parameters from recent experiment in Ref. \cite{holanda_2017}. With $\omega_s=\gamma \mu_0 h_\ssf{SSE}\simeq 0.0043\omega_M$ we can calculate the SSE torque caused line width change equals to $h_\ssf{SSE}\simeq 598\text{A/m}\approx 7.5$Oe which can be achieved \cite{holanda_2017,rezende_2014_1} for YIG/Pt system with $t_{YIG}=100$nm at $\nabla T\simeq 220$K/cm, which is realizable in experiment  \cite{holanda_2017}.

 In summary, we propose a system of magnon-polariton with EP induced by "negative damping". We show that if SSE- or SHE-STT induced "negative damping" is enough to compensate the intrinsic damping of the coupled system, the input energy is being equally distributed between subsystems. As a consequence, large photon emission from the cavity can be achieved at the EP. The thermal to microwave transformation efficiency is estimated to be about 17 percent. The induced torque or "negative damping" provides a new tool to control polariton states and to study non-Hermitian physics in magnon-polariton.	

\begin{acknowledgements}
We thank Ka Shen for fruitful and stimulating discussions. This work was financially supported by National Key Research and Development Program of China (Grant No. 2017YFA0303300) and the National Natural Science Foundation of China (No.61774017, No. 11734004, and No. 21421003).
\end{acknowledgements}
%

\end{document}